\documentclass[12pt]{iopart}

\usepackage{graphicx}

\newcommand{\bm}[1]{\mbox{\boldmath $#1$}}

\newcommand{\bms}[1]{\mbox{\scriptsize\boldmath $#1$}}

\newtheorem{proposition}{Proposition}
\newtheorem{corollary}{Corollary}

\newtheorem{theorem}{Theorem}

\begin{document}

\title[Escort Probability and Its Applications via Conformal 
Transformation]{Dually flat 
structure with escort probability and its application to 
alpha-Voronoi diagrams\footnote{Several results in this paper can be found 
in the conference paper \cite{OMA} without complete proofs.}}

\author{Atsumi Ohara$^1$, Hiroshi Matsuzoe$^2$ and Shun-ichi Amari$^3$}
\address{$^1$Department of Systems Science, 
Osaka University, Toyonaka, Osaka 560-8531, Japan}
\ead{ohara@sys.es.osaka-u.ac.jp}
\author{}
\address{$^2$Department of Computer Science and Engineering, 
Graduate School of Engineering, \\ Nagoya Institute of Technology, 
Gokiso-cho, Showa-ku, Nagoya, 466-8555 Japan}
\ead{matsuzoe@nitech.ac.jp}
\author{}
\address{$^3$Riken Brain Science Institute,
Wako-shi Hirosawa 2-1, Saitama 351-0198 Japan}
\ead{amari@brain.riken.jp}

\begin{abstract}
This paper studies geometrical structure of the manifold of
escort probability distributions 
%, which is introduced in the framework of non-extensive entropies, 
%and proves that a resultant geometry is dually
%flat in the sense of information geometry. 
and shows its new applicability to information science. 
In order to realize escort probabilities we use a 
conformal transformation that flattens so-called alpha-geometry 
of the space of discrete probability distributions, 
which well characterizes nonadditive statistics on the space.
% in the framework of affine differential geometry. 
As a result escort probabilities are proved to be flat coordinates 
of the usual probabilities for the derived dually flat structure.
%Dual pairs of potential functions and affine coordinate systems 
%on the manifold are explicitly identified, and the
%associated canonical divergence is shown to be conformal to 
%alpha-divergence.
Finally, we demonstrate that escort probabilities with 
the new structure admits a simple algorithm to compute 
Voronoi diagrams and centroids with respect to alpha-divergences.

%This note reports a conformal transformation that flattens the $\alpha$-geometry
%on the space of probability distributions. 
%In the process of this transformation we see escort probabilities play important roles.
%The resultant geometry is dually flat and 
%%, which gives us a simple geometrical interpretation for 
%the escort probabilities consist of one of the dual 
%coordinate systems for the structure.
%We discuss several properties of the obtained dually flat structure and 
%show that the associated canonical divergence is conformal 
%to the $\alpha$-divergence.
\end{abstract}

\pacs{05.90.+m, 89.70.Cf, 02.40.Hw}
%\submitto{Journal of Physics A: Mathematical and Theoretical}
\maketitle

\section{Introduction}
Escort probability is naturally induced from researches
of multifractals \cite{BS} and non-extensive statistical mechanics 
\cite{Ts09} to play an important but mysterious role.
Testing its utility in the other scientific fields would greatly help 
our understanding about it.
This motivates us to approach the escort probability by 
geometrically studying its role in information science.

%Interestingly, it shall be demonstrated that escort probability 
%also appears in a certain geometry on 
%a space of probability distributions.
%Further, it has a nice application in information science.

%The main motivation of this paper is to study another aspect of 
%the escort probability, i.e., its roles in information science, 
%in order to manage to illuminate its essence.

%Examining whether it is useful in other would be an interesting question, 
%which is related to the essence.
%physical meaning would be a direct and interesting way to 
%gain its essence. 

%We shall, however, demonstrate its new and interesting aspect 
%in information science via a geometrical approach. 

The first purpose of this paper is 
to investigate the escort probability from 
viewpoints of information geometry \cite{Amari,AN} 
and affine differential geometry \cite{NS}. 
The second is to show that escort probability with 
information geometric structure is useful 
to construction of Voronoi diagrams \cite{Ed}  
on the space of probability distributions. 

Recently, it is reported \cite{Ohara07,Ohara09} that {\em $\alpha$-geometry}, 
which is an information geometric structure 
of constant curvature, has a close relation with 
%the above areas in physics.
Tsallis statistics \cite{Ts09}.
The remarkable feature of the $\alpha$-geometry consists of 
the Fisher metric together with a one-parameter family of 
dual affine connections, called the $\alpha$-connections.
%, where alpha and - alpha connections are dual.  

%We show that the manifold of escort distributions 
%is dually flat, which is given rise to by 
%flattening the $\alpha$-geometry via a conformal transformation.

%The main feature of the $\alpha$-geometry is to consider 
%a one-parameter family of dual affine connections \cite{Amari,AN}.
%In statistics it is used to investigate geometric structure of probability 
%distributions, and recently a relation with Tsallis entropy \cite{Ts09}
%has been recognized in statistical physics 
%\cite{Ohara07,Ohara09}.
%There it is important that the $\alpha$-geometry introduces 
%a statistical manifold structure with constant curvature.

We prove that the manifold of escort probability distributions is dually flat 
by considering conformal transformations that flatten 
the $\alpha$-geometry on the manifold of
usual probability distributions.
%, and show that they 
%introduce dually flat structures.
%flattening the non-flat $\alpha$-structure.  
On the resultant manifold, escort probabilities consist of 
an affine coordinate system.  
See also \cite{OAT} for another type of flattening a curved dual manifold 
by a conformal transformation.

The result gives us a clear geometrical 
interpretation of the escort probability, and simultaneously, 
produces its new obscure links to conformality and projectivity.
Due to these two geometrical concepts, however, the obtained dually 
flat structure inherits several properties of the $\alpha$-geometry.

The dually flatness proves crucial to construction of Voronoi 
diagrams for $\alpha$-divergences, which we shall call 
$\alpha$-{\em Voronoi diagrams}.
The Voronoi diagrams on the space of probability distributions 
with the Kullback-Leibler \cite{OT,OI}, or Bregman divergences \cite{NBN}
have been recognized as important tools for various statistical modeling 
problems involving 
pattern classification, clustering, likelihood ratio test and so on. 
See also, e.g., \cite{II,Leb,Amari2} for related problems.
%in the fields of statistics, machine learning and information theory.

The largest advantage to take account of $\alpha$-divergences is 
their {\em invariance under transformations by 
sufficient statistics} \cite{Cencov} 
(See also \cite{AN} in a different viewpoint), 
which is a significant requirement for those statistical applications. 
In computational aspect, the conformal flattening of the $\alpha$-geometry 
enables us to invoke the standard algorithm \cite{ES,Ed} 
using a potential function and an upper envelop of hyperplanes 
%by lifting and projection of an upper-envelop polyhedron 
%in ${\cal E}^n \times {\bf R}$ 
with the escort probabilities as coordinates.

Section 2 is devoted to preliminaries for $\alpha$-geometry in 
the light of affine differential geometry.
In section 3, as a main result, we consider 
conformal transformations and discuss properties of the obtained 
dually flat structure.
Dual pairs of potential functions and affine coordinate systems 
on the manifold are explicitly identified, and the
associated canonical divergence is shown to be conformal to 
the $\alpha$-divergence.
Section 4 describes an application of such a flattened geometric structure 
to $\alpha$-Voronoi diagrams on the probability simplex.
The properties and a construction algorithm are discussed.
Further, a formula for $\alpha$-centroid is touched upon.

In the sequel, 
%all the affine connections are assumed torsion-free.
%$the exponent $q$ is fixed to be $(1-\alpha)/2$ 
%for $-1<\alpha<1$.
we fix the relations of two parameters $q$ and $\alpha$ as $q=(1-\alpha)/2$, 
and restrict $q>0$.

\section{Preliminaries}
We briefly introduce $\alpha$-geometry 
via affine differential geometry.
See for details \cite{Ohara07,Ohara09}.
Let $\mathcal{S}^n$ denote the $n$-dimensional probability simplex, i.e., 
\begin{equation}
	\mathcal{S}^n := \left\{ \bm{p}=(p_i) \left| \;
		p_i > 0, \; \sum_{i=1}^{n+1} p_i =1 
	\right. \right\},
\label{simpmodel}
\end{equation}
and $p_i, i=1,\cdots,n+1 $ denote
probabilities of $n+1$ states.
We introduce the $\alpha$-geometric structure on ${\cal S}^n$.
Let $\{\partial_i\}, i=1,\cdots,n$ be natural basis tangent vector fields 
on ${\cal S}^n$ defined by
\begin{equation}
	\partial_i:=\frac{\partial}{\partial p_i}
			-\frac{\partial}{\partial p_{n+1}}, \quad
			i=1,\cdots,n,
\label{basis}
\end{equation}
where $p_{n+1}=1-\sum_{i=1}^n p_i$.
Now we define a Riemannian metric $g$ on ${\cal S}^n$ 
called the {\em Fisher metric}:
\begin{eqnarray}
	g_{ij}(\bm{p})&:=& g (\partial_i, \partial_j)
=\frac{1}{p_i}\delta_{ij}+ \frac{1}{p_{n+1}} \label{Rmetric} \\
&=&\sum_{k=1}^{n+1} p_k (\partial_i \log p_k)(\partial_j \log p_k), 
\quad i,j=1,\cdots,n.
\nonumber
\end{eqnarray}
Further, define an torsion-free affine connection $\nabla^{(\alpha)}$ 
called the {\em $\alpha$-connection}, 
which is represented in its coefficients by
\begin{equation}
	 \Gamma^{(\alpha)k}_{ij}(\bm{p}) 
%:= \sum_{l=1}^{n} g^{kl} \Gamma^{(\alpha)}_{ijl}
%\tilde g(\tilde \nabla^{(\alpha)}_{\partial_i}\partial_j, \partial_l) 
%=\frac{1+\alpha}{2}
%\left(- \frac{1}{p_i}\delta_{ij}^k + \frac{p_k}{p_i}\delta_{ij}+
%\frac{p_k}{p_{n+1}}\right)
=\frac{1+\alpha}{2}
\left(-\frac{1}{p_k}\delta_{ij}^k + p_k g_{ij}\right), 
\quad i,j,k=1,\cdots,n,
\label{alpha}
\end{equation}
where $\delta_{ij}^k$ is equal to one if $i=j=k$ and zero otherwise.
Then we have the $\alpha$-covariant derivative $\nabla^{(\alpha)}$,
which gives
\[
	\nabla^{(\alpha)}_{\partial_i} \partial_j
		=\sum_{k=1}^{n} \Gamma^{(\alpha)k}_{ij} \partial_k,
\]
when it is applied to the vector fields $\partial_i$ and $\partial_j$.

There are two specific features for the 
$\alpha$-geometry on ${\cal S}^n$ defined in such a way.
First, the triple $({\cal S}^n,g,\nabla^{(\alpha)})$ is a
{\em statistical manifold} \cite{Lau} (See appendix A for its definition), 
i.e., we can confirm that the following relation holds:
\begin{equation}
	Xg(Y,Z)=g(\nabla^{(\alpha)}_X Y,Z)+g(Y,\nabla^{(-\alpha)}_X Z),
	\quad X,Y,Z \in {\cal X}({\cal S}^n),
\label{dc}
\end{equation}
where ${\cal X}({\cal S}^n)$ denotes the set of all tangent vector fields 
on ${\cal S}^n$.
Two statistical manifolds $({\cal S}^n,g,\nabla^{(\alpha)})$ and 
$({\cal S}^n,g,\nabla^{(-\alpha)})$ are said {\em mutually dual}.

The other is that $({\cal S}^n,g,\nabla^{(\alpha)})$ is a manifold of 
constant curvature $\kappa=(1-\alpha^2)/4$, i.e.,
\[
	 R^{(\alpha)}(X,Y)Z= \kappa \{g(Y,Z)X - g(X,Z)Y \},
\]
where $R^{(\alpha)}$ is the curvature tensor 
with respect to $\nabla^{(\alpha)}$.
From this property the well-known nonadditive formula 
of the Tsallis entropy can be derived \cite{Ohara07}.

In \cite{Ohara09} we have discussed the $\alpha$-geometry on ${\cal S}^n$ 
from a viewpoint of {\em affine differential geometry} \cite{NS}.
Consider the immersion $f$ of ${\cal S}^n$ into 
${\bf R}^{n+1}_+$ by
\begin{equation}
	f: \bm{p}=(p_i) \mapsto \bm{x}=(x^i)=(L^{(\alpha)}(p_i)), 
	\quad i=1,\cdots,n+1,
\label{immersion}
\end{equation}
where $(x^i), i=1,\cdots,n+1$ is 
the canonical flat coordinate system of ${\bf R}^{n+1}$ 
and the function $L^{(\alpha)}$ is defined by
\[
	L^{(\alpha)}(t) :=\frac{2}{1-\alpha} t^{(1-\alpha)/2}=\frac{1}{q}t^q.
\]
Note that $f({\cal S}^n)$ is a level hypersurface in 
the ambient space ${\bf R}^{n+1}_+$ 
represented by $\Psi(\bm{x})=2/(1+\alpha)$, where 
%$\psi$ is defined by
\begin{equation}
 \Psi(\bm{x}) := \frac{2}{\alpha +1} \sum_{i=1}^{n+1}
 \left(\frac{1-\alpha}{2} x^i \right)^{2/(1-\alpha)}
=\frac{1}{1-q} \sum_{i=1}^{n+1}
 \left(q x^i \right)^{1/q}.
%\qquad \alpha \not= \pm 1.
\label{potential}
\end{equation}
We choose a transversal vector $\xi$ on the level hypersurface by
\begin{equation}
	\xi:=  \sum_{i=1}^{n+1} \xi^i \frac{\partial}{\partial x^i},
	\quad \xi^i= -q(1-q) x^i = -\kappa x^i.	
\label{trans}
\end{equation}
Then we can confirm that the affine immersion $(f,\xi)$ realizes 
the $\alpha$-geometry on ${\cal S}^n$ \cite{Ohara09}.
Hence, it would be possible to develop theory of 
the $\alpha$-geometry and Tsallis statistics 
with ideas of affine differential geometry \cite{MTA}.

Further, the {\em escort probability} \cite{BS} naturally appears 
in this setup.
The escort probability $\bm{P}=(P_i)$ associated with $\bm{p}=(p_i)$ is 
the normalized version of $(p_i)^q$, and is defined by
\begin{equation}
\fl	P_i(\bm{p}):=\frac{(p_i)^q}{\sum_{j=1}^{n+1} (p_j)^q }
	=\frac{x^i}{Z_q}, \; i=1,\cdots,n+1,
	\quad Z_q(\bm{p}):=\sum_{i=1}^{n+1} x^i(\bm{p}), 
	\; \bm{x}(\bm{p}) \in f({\cal S}^n).
%	=\frac{L^{(\alpha)}(p_i)}{Z_q},
%	\quad Z_q:=\sum_{i=1}^{n+1} L^{(\alpha)}(p_i),
\label{projective}
\end{equation}
Hence, the simplex ${\cal E}^n$ in the ambient space ${\bf R}^{n+1}_+$, 
i.e.,
\[
	{\cal E}^n :=\left\{ \bm{x}=(x^i) \left| 
	\sum_{i=1}^{n+1} x^i =1, \; x^i >0 \right. \right\}
\]
represents the set of escort distributions $\bm{P}$. 
%(See the figure 1 in \cite{Ohara09}).

Note that the element $\bm{x}^*=(x_i^*)$ in the dual space of 
${\bf R}^{n+1}$ defined by
\[
	x^*_i(\bm{p}):=L^{(-\alpha)}(p_i)=\frac{1}{1-q}(p_i)^{1-q}, 
	\quad i=1,\cdots,n+1,
\]
meets 
\[
	x^*_i (\bm{p}) = \frac{\partial \Psi}{\partial x^i}(\bm{x}(\bm{p})).
\]
Hence, it satisfies \cite{Ohara09} 
\begin{equation}
	-\sum_{i=1}^{n+1} \xi^i(\bm{p}) x^*_i(\bm{p}) =1, \quad
	\sum_{i=1}^{n+1} x^*_i(\bm{p}) X^i =0,
\end{equation}
for an arbitrary vector $X= \sum_{i=1}^{n+1} X^i \partial /\partial x^i$ 
at $\bm{x}(\bm{p})$ tangent to $f({\cal S}^n)$.
Thus, $-\bm{x}^*(\bm{p})$ can be interpreted as 
the {\em conormal map} \cite{NS}.

\section{A conformally and projectively flat geometric structure and escort probabilities}
In this section we show a main result.
For this purpose, we consider a conformal and projective transformation 
\cite{Kurose90,Kurose94,Kurose02,Matsu}
of the $\alpha$-geometry to introduce a dually flat one.
This flattening of the $\alpha$-geometry conserves 
some of its properties.
The escort probabilities $(P_i)$ are found to represent 
one of mutually dual affine coordinate systems in the induced geometry.
While the many functions or geometric quantities introduced in this section 
depend on the parameter $\alpha$ or $q$, we omit them for the brevity.

Let us define a function $\lambda$ on ${\cal S}^n$ by
\[
	\lambda (\bm{p}):=\frac{1}{Z_q}
	=\frac{1}{\sum_{i=1}^{n+1} L^{(\alpha)}(p_i)},
\]
which depends on $\alpha$.
Then, from (\ref{projective}) ${\cal E}^n$ is regarded as 
the image of ${\cal S}^n$ for another immersion $\tilde f:=\lambda f$, i.e.,
\[
	\tilde f: {\cal S}^n \ni (p_i) 
		\mapsto (P_i) \in {\cal E}^n,
	\quad i=1,\cdots,n+1,
\]
and $(P_1, \cdots, P_n)$ is interpreted as another coordinate system 
of ${\cal S}^n$.
Note that the inverse mapping $\tilde f ^{-1}$ is well-defined by
\[
	\tilde f ^{-1}: (P_i) \mapsto (p_i) = 
	\left( \frac{(P_i)^{1/q}}{\sum_{j=1}^{n+1} (P_j)^{1/q}} \right),
	\quad i=1,\cdots,n+1.
\]
It would be a natural way to introduce geometric structure on ${\cal E}^n$ 
(and hence on ${\cal S}^n$) via the affine immersion 
$(\tilde f, \tilde \xi)$ by taking a suitable 
transversal vector $\tilde \xi$, 
similarly to the case of the $\alpha$-geometry mentioned above. 
Since ${\cal E}^n$ is a part of a hyperplane in ${\bf R}^{n+1}$, 
the canonical affine connection of ${\bf R}^{n+1}$ induces 
a flat connection, denoted by $D^{({\rm E})}$, on ${\cal E}^n$.
However, for the same reason, we cannot define a Riemannian metric 
in this way\footnote{In affine differential geometry, 
a Riemannian metric is realized as the affine fundamental form 
of an affine immersion \cite{NS}.} because it vanishes on ${\cal E}^n$, 
regardless of any choice of the transversal vector $\tilde \xi$.
%This is because ${\cal E}^n$ is a part of the hyperplane in ${\bf R}^{n+1}$.

The idea we adopt here is to define a Riemannian metric by utilizing a 
property of $({\cal S}^n,g,\nabla^{(\alpha)})$ called 
{\em $-1$-conformal flatness}.
Based on the results proved by Kurose \cite{Kurose90, Kurose94}, 
we conclude that the manifold $({\cal S}^n,g,\nabla^{(\alpha)})$ is 
$\pm 1$-conformally flat (See Appendix A for its definition) 
because it is a statistical manifold of constant curvature.

%Hence the escort probability $P_i, i=1,\cdots,n$ are nothing but 
%the canonical flat coordinates 
%\footnote{Exactly speaking, so are $P_i, i=1,\cdots,n$ because 
%$P_{n+1}=1-\sum_{i=1}^n P_i$.}
%of the manifold $({\cal E}^n, h, D^{(E)})$,
%which is $-1$-conformally equivalent to 
%the manifold $({\cal S}^n,g,\nabla^{(\alpha)})$.
%XXXX This is not a strictly exact statement!

Actually, let $\nabla^*$ be the flat connection\footnote{For the sake 
of notational consistency with the existing literature, 
e.g., \cite{Amari,AN}, we first define $\nabla^*$, 
and later $\nabla$ as the dual of $\nabla^*$.} 
on ${\cal S}^n$ defined 
%with the differential $\tilde f_*$ by
with $D^{({\rm E})}$ and the differential $\tilde f_*$ by
\[
	\tilde{f}_*(\nabla^*_X Y)=D^{({\rm E})}_{\tilde{f}_* X} \tilde{f}_* Y,
	\quad X, Y \in {\cal X}({\cal S}^n).
\]
Then, we can prove that $\nabla^{(\alpha)}$ and $\nabla^*$ are 
{\em projectively equivalent} \cite{NS}, i.e., it holds that
\begin{equation}
\nabla^*_X Y = \nabla^{(\alpha)}_X Y +d(\ln \lambda) (Y)X+d (\ln \lambda)(X)Y,
\quad X, Y \in {\cal X}({\cal S}^n).
\label{proj_eq}
\end{equation}
Hence, if we define another Riemannian metric $h$ on ${\cal S}^n$ by
\begin{equation}
	h(X,Y):= \lambda g(X,Y),	\quad X, Y \in {\cal X}({\cal S}^n),
\end{equation}
then, $({\cal S}^n,g,\nabla^{(\alpha)})$ is $-1$-conformally equivalent 
to $({\cal S}^n,h,\nabla^*)$ equipped with a flat connection $\nabla^*$.
Further, the manifold $({\cal S}^n,h,\nabla^*)$ can be proved to be a 
statistical manifold (See Appendix B).

Using the conormal map $-\bm{x}^*(\bm{p})$, 
we can define the {\em $\alpha$-divergence} as a 
{\em contrast function} (See Appendix A) 
%\footnote{A divergence that is 
%compatible with statistical manifold structure 
%in the sense of Eguchi \cite{Eguchi} is called a {\em contrast function}.} 
inducing $(g,\nabla^{(\alpha)},\nabla^{(-\alpha)})$ as follows \cite{Kurose94}:
\begin{eqnarray*}
	D^{(\alpha)}(\bm{p},\bm{r})
	&=&-\sum_{i=1}^{n+1} x^*_i(\bm{r})(x^i(\bm{p}) 
	-x^i(\bm{r})) \\
	&=&\langle - \bm{x}^*(\bm{r}), \bm{x}(\bm{p}) 
	-\bm{x}(\bm{r}) \rangle 
	=\frac{1}{\kappa} 
	- \langle \bm{x}^*(\bm{r}), \bm{x}(\bm{p}) \rangle.
\end{eqnarray*}

The statistical manifolds $({\cal S}^n,g,\nabla^{(-\alpha)})$ and 
$({\cal S}^n,g,\nabla^{(\alpha)})$ are dual in the sense of (\ref{dc}).
Further, it is known \cite{AN} that 
there exists the unique affine flat connection $\nabla$ on ${\cal S}^n$, 
dual with respect to $(h,\nabla^*)$.
%Then, $({\cal S}^n,g,\nabla^{(-\alpha)})$ is proved 
%1-conformally equivalent to $({\cal S}^n,h,\nabla)$ \cite{Kurose94}.
%This fact directly implies \cite{Kurose94} that a contrast function 
Then, according to \cite{Kurose94}, it is proved that 
$({\cal S}^n,h,\nabla)$ is 1-conformally equivalent to 
$({\cal S}^n,g,\nabla^{(-\alpha)})$ and 
a contrast function $\rho$ inducing $(h,\nabla,\nabla^*)$ is given by scaling 
$D^{(-\alpha)}$ (See Appendix A)
%, which is a 
%contrast function inducing $(g,\nabla^{(-\alpha)},\nabla^{(\alpha)})$, 
as follows:
\begin{eqnarray}
	\rho (\bm{p},\bm{r})
	&=& \lambda({\bm{r}}) D^{(-\alpha)}(\bm{p},\bm{r}) 
%	= e^{\log \lambda(q)} \rho(p,q)
	=\frac{1}{Z_q(\bm{r})} D^{(-\alpha)}(\bm{p},\bm{r}) \nonumber \\
	&=& \frac{1}{Z_q(\bm{r})} 
	\langle -\bm{x}(\bm{r}), \bm{x}^*(\bm{p}) 
	-\bm{x}^*(\bm{r}) \rangle 
	= \langle -\bm{P}(\bm{r}), \bm{x}^*(\bm{p}) 
	-\bm{x}^*(\bm{r}) \rangle.
%	\; \lambda=\frac{1}{Z_q}.
\label{conf_div}
\end{eqnarray}
We shall call $\rho$ a {\em conformal divergence}.

%\noindent
%XXXXXXXX  This is the same one introduced in Amari memo 09/02/20 XXXXXXXX

Now, since $({\cal S}^n,h,\nabla,\nabla^*)$ is a 
dually flat space, the standard result in \cite{Amari,AN} suggests 
that there exist {\em mutually dual affine coordinate systems} 
$(\theta^1,\cdots,\theta^n)$ and $(\eta_1,\cdots,\eta_n)$, 
a {\em potential function} $\psi(\bm{\theta})$ and {\em its conjugate} 
$\psi^*(\bm{\eta})$ satisfying 
\begin{equation}
	\eta_i = \frac{\partial \psi}{\partial \theta^i}, 
	\quad \theta^i=\frac{\partial \psi^*}{\partial \eta_i}, \quad
	i=1,\cdots,n.
\label{first_der}
\end{equation}
They completely determine dually flat structure, i.e.,  
the coefficients of $h$, $\nabla$ and $\nabla^*$ are derived as 
the second and third derivatives of $\psi$ or $\psi^*$, for example,
\begin{eqnarray*}
&	\displaystyle
	h_{ij}=h\left(\frac{\partial}{\partial \theta^i}, 
	\frac{\partial}{\partial \theta^j} \right)
	=\frac{\partial^2 \psi}{\partial \theta^i \partial \theta^j}, \quad
	h^{ij}=h\left(\frac{\partial}{\partial \eta_i}, 
	\frac{\partial}{\partial \eta_j} \right)
	=\frac{\partial^2 \psi^*}{\partial \eta_i \partial \eta_j},& \\
&	\displaystyle
	\Gamma_{ijk}=h \left(\nabla_{\frac{\partial}{\partial \theta^i}} 
	\frac{\partial}{\partial \theta^j}, \frac{\partial}{\partial \theta^k} 
	\right)=0, \quad
	\Gamma^*_{ijk}=h \left(\nabla^*_{\frac{\partial}{\partial \theta^i}} 
	\frac{\partial}{\partial \theta^j}, \frac{\partial}{\partial \theta^k} 
	\right)=\frac{\partial^3 \psi}{\partial \theta^i \partial \theta^j 
	\partial \theta^k}, &
\end{eqnarray*}
and so on.
In order to identify $\psi, \psi^*, \theta^i$ and $\eta_i$ explicitly 
{\em without integrating $h_{ij}$ or $h^{ij}$}, we shall search for them 
by examining whether the conformal divergence $\rho$ can be represented 
in the form of the {\it canonical divergence} \cite{AN}, i.e., 
\begin{equation}
	\rho (\bm{p},\bm{r})
	=\psi(\bm{\theta}(\bm{p}))+\psi^*(\bm{\eta}(\bm{r}))
	-\sum_{i=1}^n \theta^i(\bm{p}) \eta_i(\bm{r}).
\label{can_div}
\end{equation}
with the constraints (\ref{first_der}).
If this is possible, we can directly prove 
from (\ref{cont1}) and (\ref{cont2}) that the obtained 
$\psi, \psi^*, (\theta^1,\cdots,\theta^n)$ and $(\eta_1,\cdots,\eta_n)$ 
%in (\ref{can_div}) 
are pairs of dual potential functions and affine coordinate systems 
associated with $({\cal S}^n,h,\nabla,\nabla^*)$.

%An easy way to identify pairs of potential functions and 
%dual coordinate systems without integrating $h$ is XXXX
%Further, for a dually flat space, the compatible contrast function 
%must be written uniquely 
%From (\ref{conf_div}) and (\ref{can_div}), 
%From these two facts, we will derive explicit representations of 
%$\psi, \psi^*, \theta^i$ and $\eta_i$.
Before showing the result, we define, for $0<q$ with $q \not= 1$, 
two functions by 
\[
	\ln_q (s):= \frac{s^{1-q}-1}{1-q}, \; s \ge 0, \quad
	\exp_q (t):=[1+(1-q)t]_+^{1/(1-q)}, \; t \in {\bf R},
\]
where $[t]_+:=\max\{0,t \}$, and the so-called Tsallis entropy \cite{Ts88} by 
\[
	S_q(\bm{p}):= \frac{\sum_{i=1}^{n+1} (p_i)^q -1}{1-q}.
\]
Note that $s=\exp_q (\ln_q (s))$ holds and 
they respectively recover the usual 
logarithmic, exponential function 
and the Boltzmann-Gibbs-Shannon entropy $-\sum_{i=1}^{n+1} p_i \ln p_i$ 
when $q \rightarrow 1$.
For $q>0$, $\ln_q(s)$ is concave on $s>0$.

\begin{theorem}
\label{thm1}
For the dually flat space $({\cal S}^n,h,\nabla,\nabla^*)$ defined via 
$\pm 1$-conformal transformation from 
$({\cal S}^n,g,\nabla^{(\alpha)},\nabla^{(-\alpha)})$, 
the associated potential functions $\psi, \psi^*$, and 
dually flat affine coordinate systems $(\theta^1,\cdots,\theta^n)$ and 
$(\eta_1,\cdots,\eta_n)$ are represented as follows:
%\[
%	\theta^i(\bm{p}) = x^*_i(\bm{p})-x^*_{n+1}(\bm{p}),
%	\eta_i (\bm{p}) = P_i(\bm{p}), 
%	\quad i=1,\cdots,n  \\ 
%\]
\begin{eqnarray*}
	\theta^i(\bm{p}) &=& x^*_i(\bm{p})-x^*_{n+1}(\bm{p}), 
	\quad i=1,\cdots,n  \\ 
	\eta_i (\bm{p}) &=& P_i(\bm{p}), 
	\quad i=1,\cdots,n  \\ 
	\psi(\bm{\theta}(\bm{p})) &=& -\ln_q (p_{n+1}), \\
	\displaystyle \psi^*(\bm{\eta}(\bm{p})) &=& \frac{1}{\kappa} 
				\left(\lambda(\bm{p}) -q \right)
	=\frac{1}{1-q} \left(\sum_{i=1}^{n+1} (\eta_i)^{1/q} \right)^q
	-\frac{1}{1-q},  
%		=-x^*_{n+1}(\bm{p})+\frac{1}{1-q},
\end{eqnarray*}
where $\kappa=(1-\alpha^2)/4=q(1-q)$ 
is the scalar curvature of 
$({\cal S}^n,g,\nabla^{(\alpha)},\nabla^{(-\alpha)})$ and 
$\eta_{n+1}:=P_{n+1}(\bm{p})=1-\sum_{i=1}^n P_i(\bm{p})$.
Further, the coordinate systems $(\theta^1,\cdots,\theta^n)$ and 
$(\eta_1,\cdots,\eta_n)$ are $\nabla$- and $\nabla^*$-affine, respectively.
\end{theorem}
\noindent
Proof) As is mentioned above we have only to check that 
%the canonical divergence constructed by (\ref{can_div}) with 
the potential functions $\psi, \psi^*$ and dual affine coordinates 
$\theta^i, \eta_i$ in the statement satisfy 
(\ref{first_der}) and (\ref{can_div}) for the conformal divergence $\rho$.
% of $({\cal S}^n,h,\nabla)$. 
%Using the contrast function $\rho$ of $({\cal S}^n,h,\nabla)$ 
%in (\ref{conf_div}), the contrast function 
%$\rho^*$ of $({\cal S}^n,h,\nabla^*)$ is given by
%\[
%	\rho(\bm{p},\bm{r})=\rho^*(\bm{r},\bm{p}).
%\]
First, substitute them directly 
%the potential functions and dual affine coordinates in the statement 
to the right-hand side of (\ref{can_div}) and modify it 
caring for the relation $\eta_{n+1}=1-\sum_{i=1}^n \eta_i$, 
then we see that it coincides with $\rho(\bm{p},\bm{r})$ in (\ref{conf_div}).
Next, since it holds that $\ln_q (p_i)=x^*_i(\bm{p})-1/(1-q)$,
we can alternatively represent
\[
	\theta^i(\bm{p})=\ln_q (p_i) - \ln_q (p_{n+1})
	=\ln_q (p_i) + \psi(\bm{\theta}(\bm{p})), \quad i=1,\cdots,n.
\]
Hence, for $\theta^{n+1}\equiv 0$ it holds 
\[
	1= \sum_{i=1}^{n+1} p_i = \sum_{i=1}^{n+1} \exp_q (\theta^i - \psi).
\]
Differentiating the both sides by $\theta^j, j=1,\cdots,n$, we have
\[
	0= \sum_{i=1}^{n+1} \left(\delta_{ij}
	-\frac{\partial \psi}{\partial \theta^j}\right) (p_i)^q
	=(p_j)^q - \frac{\partial \psi}{\partial \theta^j} 
	\sum_{i=1}^{n+1}(p_i)^q, \qquad j=1,\cdots,n.
\]
Thus, the left equation of (\ref{first_der}) holds. 
Finally, note that the conformal factor is represented by
\begin{equation}
	\lambda(\bm{p})=\frac{1}{Z_q(\bm{p})}
	=\frac{q}{\sum_{i=1}^{n+1} (p_i)^q}
	=\frac{q}{(\exp_q (S_q(\bm{p})))^{1-q}}.
\label{c_fctr1}
\end{equation}
Using the formula \cite{SW}:
\[
	\exp_q(S_q(\bm{p})) 
	= \exp_{\frac{1}{q}} \left(S_{\frac{1}{q}}(\bm{P}) \right),
\]
we see that
\[
	\lambda(\bm{p})=
	q \left( \exp_{\frac{1}{q}} \left(S_{\frac{1}{q}}(\bm{P}) \right) 
	\right)^{q-1}
	=q \left(\sum_{i=1}^{n+1}(P_i)^{\frac{1}{q}} \right)^q.
\]
Hence, the second equality in the expression of $\psi^*$ holds. 
The right equation of (\ref{first_der}) follows if you again 
recall $\eta_{n+1}=1-\sum_{i=1}^n \eta_i$.
\hfill Q.E.D.
\begin{corollary}
The escort probabilities $P_i, i=1,\cdots,n$ are canonical 
affine coordinates of the flat affine connection $\nabla^*$ on ${\cal S}^n$.
\end{corollary}

%There are three remarks. 
%First, 
{\it Remark 1}: 
Since the conformal factor $\lambda$ in (\ref{c_fctr1})
can be alternatively represented by
\[
	\lambda(\bm{p})
%=\frac{1}{Z_q(\bm{p})}
%=\frac{q}{\sum_{i=1}^{n+1} (p_i)^q}
	=\frac{q}{(\exp_q (S_q(\bm{p})))^{1-q}}
	=\kappa \ln_q \left( \frac{1}{\exp_q(S_q(\bm{p}))} \right)+q,
\]
we have another expression of $\psi^*$, i.e,
\[
	\psi^*= \ln_q \left( \frac{1}{\exp_q(S_q(\bm{p}))} \right).
\]
Thus, the potentials and dual coordinates given in the proposition 
recover the standard ones \cite{Amari,AN} when $q \rightarrow 1$, i.e, 
\[ \fl
	\psi \rightarrow -\ln p_{n+1}, \quad
	\psi^* \rightarrow \sum_{i=1}^{n+1} p_i \log p_i \quad
	\theta^i \rightarrow \log (p_i/p_{n+1}), \quad
	\eta_i \rightarrow p_i, \quad i=1,\cdots,n.
\]
Note that $-\psi^*$ coincides with the entropy studied in \cite{LV,RA,WS} 
and referred to as the {\em normalized Tsallis entropy}.
The conformal (or scaling) factor $\lambda$ often 
appears in the study of the $q$-analysis. 

%Next, 
{\it Remark 2}:
Similarly to the above conformal transformation of 
$({\cal S}^n,g,\nabla^{(\alpha)})$, we can define 
another one for $({\cal S}^n,g,\nabla^{(-\alpha)})$ with
a conformal factor
\[
	\lambda'(\bm{p}):=\frac{1}{\sum_{i=1}^{n+1} L^{(-\alpha)}(p_i)},
\]
and construct another dually flat structure 
$(h'=\lambda'g,\nabla',\nabla'^*)$.
Hence, the following relations among them hold (See Figure \ref{fig1}).
\begin{figure}[h]
\[ \fl
\begin{array}{ccccc}
&({\cal S}^n,h',\nabla') & \stackrel{\rm dual}{\longleftrightarrow} & 
({\cal S}^n,h',\nabla'^*) & \\
1 \mbox{-conformally equivalent} & \updownarrow && \updownarrow & -1 \mbox{-conformally equivalent} \\
& ({\cal S}^n,g,\nabla^{(\alpha)}) & \stackrel{\rm dual}{\longleftrightarrow} & ({\cal S}^n,g,\nabla^{(-\alpha)}) & \\
 -1 \mbox{-conformally equivalent} & \updownarrow && \updownarrow & 1 \mbox{-conformally equivalent} \\
& ({\cal S}^n,h,\nabla^*) & \stackrel{\rm dual}{\longleftrightarrow} & 
({\cal S}^n,h,\nabla) &
\end{array}
\]
\caption{Relations among geometries}
\label{fig1}
\end{figure}

%Finally, 
{\it Remark 3}: 
Because of the projective equivalence (\ref{proj_eq}), 
a submanifold in ${\cal S}^n$ is $\nabla^{(\alpha)}$-autoparallel 
if and only if it is $\nabla^*$-autoparallel.
In particular, the set of distributions constrained 
with the normalized $q$-expectations (escort averages) \cite{Ts09} is 
a simultaneously $\nabla^{(\alpha)}$- and $\nabla^*$-autoparallel 
submanifold in ${\cal S}^n$.

%\begin{remark}
%\[
%	L[p(x)]:=\frac{p(x)^q}{\sum_{x=1}^{n+1} p(x)^q}, 
%	\quad L^*[p(x)]:=\ln_q p(x)
%\]
%\end{remark}

\section{Applications to construction of alpha-Voronoi diagrams 
and alpha-centroids}

For given $m$ points $\bm{p}_1, \cdots, \bm{p}_m$ on ${\cal S}^n$ 
we define {\em $\alpha$-Voronoi regions} on ${\cal S}^n$ using 
the $\alpha$-divergence as follows:
\[
	{\rm Vor}^{(\alpha)}(\bm{p}_k):=\bigcap_{l \not= k} 
	\{ \bm{p} \in {\cal S}^n | 
	D^{(\alpha)}(\bm{p},\bm{p}_k)< D^{(\alpha)}(\bm{p},\bm{p}_l) \},
	\quad k=1,\cdots,m.
\]
An {\em $\alpha$-Voronoi diagram} on ${\cal S}^n$ is a collection of the 
$\alpha$-Voronoi regions and their boundaries.
Note that $D^{(\alpha)}$ approaches the Kullback-Leibler divergence 
if $\alpha \rightarrow -1$, and $D^{(0)}$ is called 
the Hellinger distance.
If we use the {\em R\'enyi divergence of order} $\alpha \not=1$ \cite{Renyi} 
defined by
\[
	D_\alpha(\bm{p},\bm{r}):= \frac{1}{\alpha-1} \ln 
		\sum_{i=1}^{n+1} (p_i)^{\alpha} (r_i)^{1-\alpha},
\]
instead of the $\alpha$-divergence, ${\rm Vor}^{(1-2\alpha)}(\bm{p}_k)$ 
gives the corresponding Voronoi region because of their one-to-one 
functional relationship.

The standard algorithm 
using projection of a polyhedron \cite{ES,Ed} commonly works well 
to construct Voronoi diagrams for 
the Euclidean distance \cite{Ed}, 
the Kullback-Leibler \cite{OI} and Bregman divergences \cite{NBN},
respectively.
The algorithm is applicable 
if a distance function is represented by 
the remainder of the first order Taylor expansion of 
a convex potential function in a suitable coordinate system.
Geometrically speaking, this is satisfied if i) the divergence is a 
canonical one for a certain dually flat structure and 
ii) its affine coordinate system is chosen 
to realize the corresponding Voronoi diagrams.
In this coordinate system with one extra complementary coordinate 
the polyhedron is expressed as the upper envelop of 
%an arrangement composed by 
$m$ hyperplanes tangent to the potential function.
%See the cases of the Euclidean distance \cite{Ed}, 
%the Kullback-Leibler \cite{OI} and Bregman divergences \cite{NBN},
%respectively.

A problem for the case of the $\alpha$-Voronoi diagram is 
that the $\alpha$-divergence 
on ${\cal S}^n$ {\em cannot} be represented as a remainder of any 
convex potentials.
The following theorem, however, claims that the problem is resolved by 
conformally transforming the $\alpha$-geometry 
to the dually flat structure $(h,\nabla,\nabla^*)$ and 
using the conformal divergence $\rho$ and escort probabilities as a 
coordinate system. 
%(See the figures 2 and 3.)

\begin{figure}[bhtp]
\vspace{0cm}
\begin{center}
\includegraphics[width=15cm]{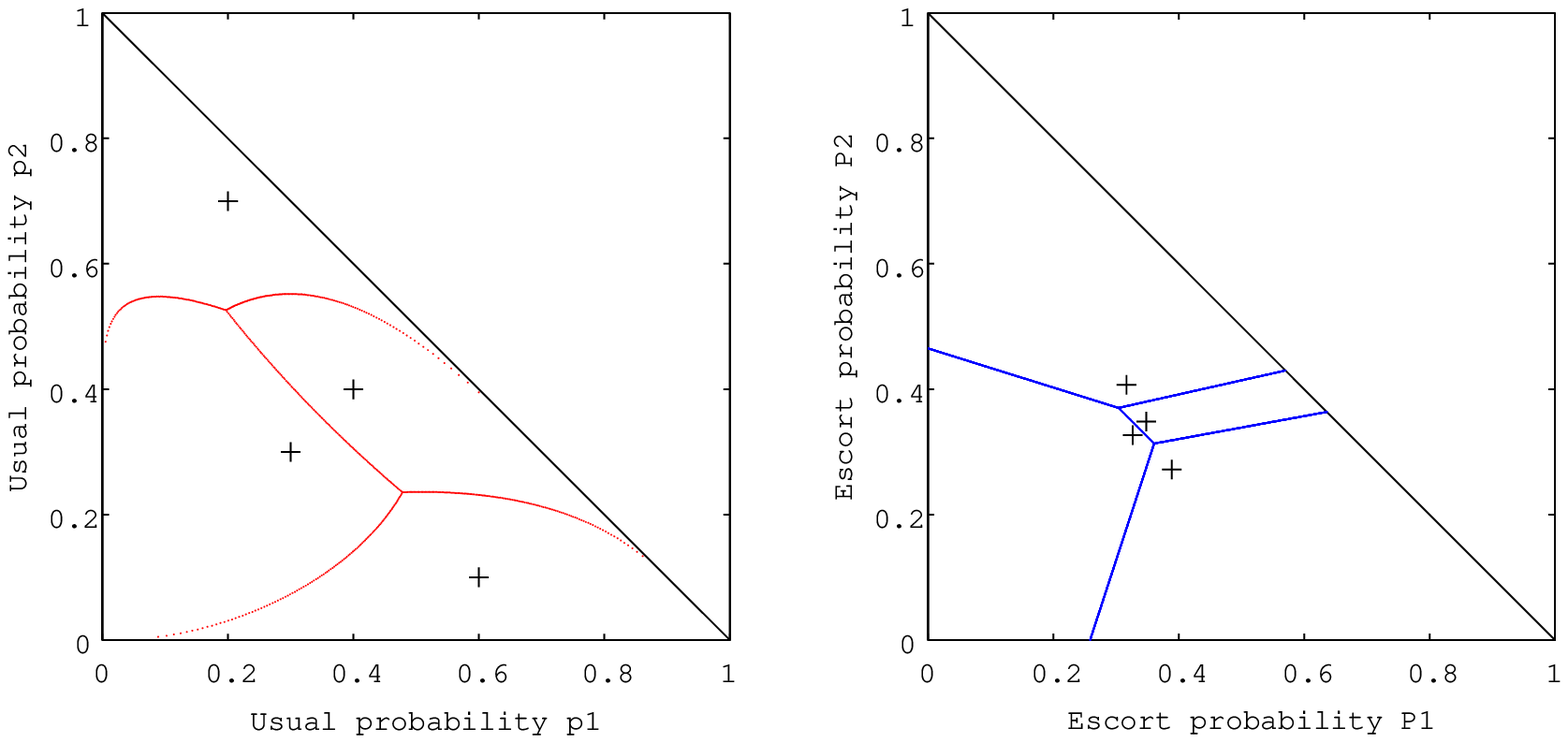}
\caption{An example of $\alpha$-Voronoi diagram on ${\cal S}^2$ (left) 
for $\alpha=0.6$ (or $q=0.2$) and the corresponding one on ${\cal E}^2$ 
(right).} 
\includegraphics[width=15cm]{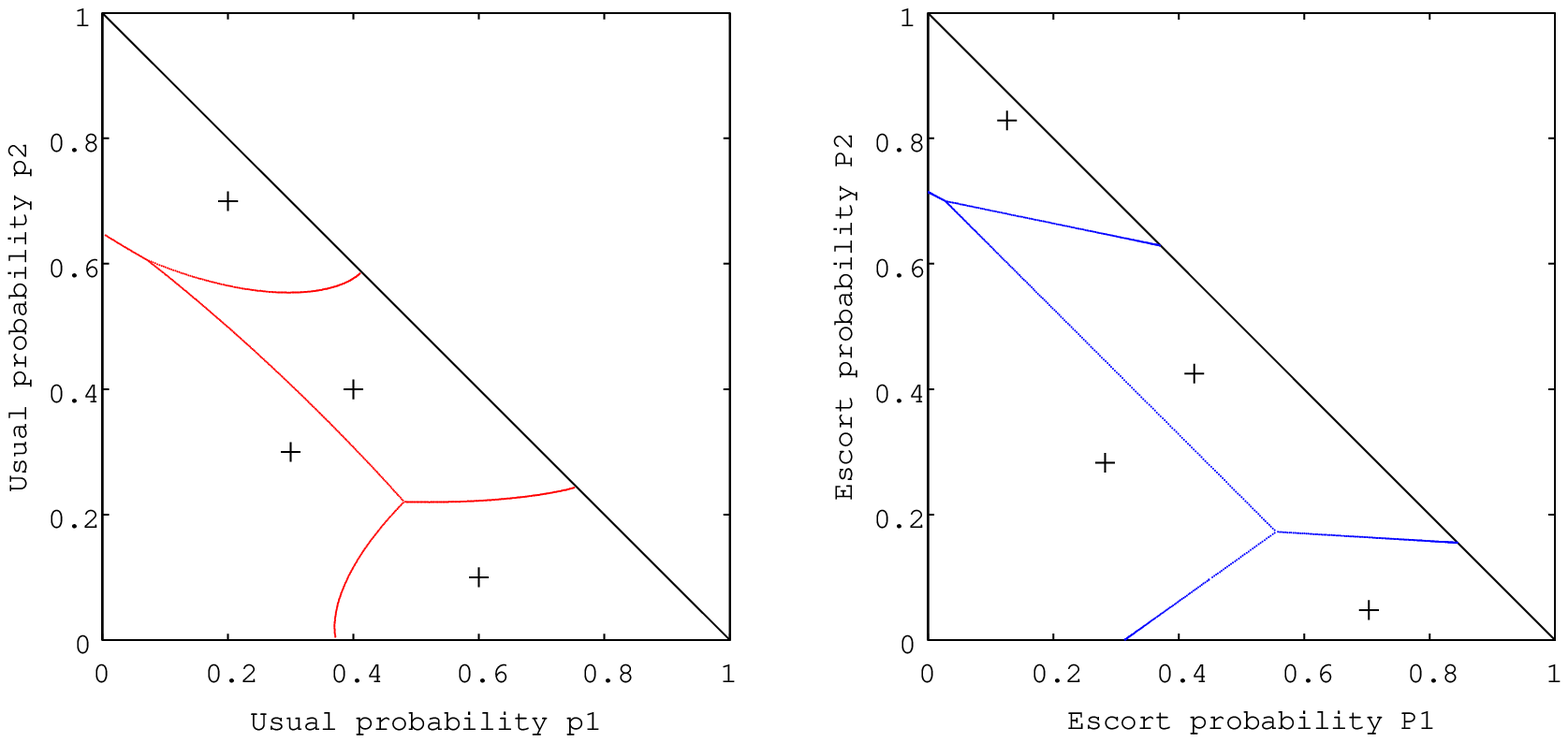}
\caption{An example of $\alpha$-Voronoi diagram on ${\cal S}^2$ (left) 
for $\alpha=-2$ (or $q=1.5$) and the corresponding one on ${\cal E}^2$ 
(right).} 
\end{center}
\end{figure}

Here, we denote the point on ${\cal E}^n$ by $\bm{P}=(P_1,\cdots,P_n)$ because 
$P_{n+1}=1-\sum_{i=1}^n P_i$.

\begin{theorem}
\begin{description}
\item{i)} The bisector of $\bm{p}_k$ and $\bm{p}_l$ defined by 
$\{\bm{p} |D^{(\alpha)}(\bm{p},\bm{p}_k)=D^{(\alpha)}(\bm{p},\bm{p}_l)\}$ 
is a simultaneously $\nabla^{(\alpha)}$- and $\nabla^*$-autoparallel 
hypersurface on ${\cal S}^n$.
%which is represented as  hyperplane on ${\cal E}^n$.
\item{ii)} Let ${\cal H}_k, k=1,\cdots,m$ be the hyperplane in 
${\cal E}^n \times {\bf R}$ which is respectively tangent 
at $(\bm{P}_k,  \psi^*(\bm{P}_k))$ to the hypersurface 
$\{(\bm{P},y) | y=\psi^*(\bm{P}) \}$, where 
$\bm{P}_k=\bm{P}(\bm{p}_k)$.
The $\alpha$-Voronoi diagram 
can be constructed on ${\cal E}^n$ as the projection of 
the upper envelope of ${\cal H}_k$'s along the $y$-axis.
%the arrangement in ${\cal E}^n \times {\bf R}$ composed by .
\end{description}
\end{theorem}
\noindent
Proof) i) Consider the $\nabla^{(-\alpha)}$-geodesic $\gamma^{(-\alpha)}$ 
connecting $\bm{p}_k$ and $\bm{p}_l$, and 
let $\bar{\bm{p}}$ be the midpoint on $\gamma^{(-\alpha)}$ satisfying 
$D^{(\alpha)}(\bar{\bm{p}},\bm{p}_k)=D^{(\alpha)}(\bar{\bm{p}},\bm{p}_l)$.
Denote by ${\cal B}$ the $\nabla^{(\alpha)}$-autoparallel hypersurface 
that is orthogonal to $\gamma^{(-\alpha)}$ and contains $\bar{\bm{p}}$.
Then, for all $\bm{r} \in {\cal B}$, the modified Pythagorean theorem 
\cite{Kurose94,Ohara07} implies the following equality:
\begin{eqnarray*}
\fl 	D^{(\alpha)}(\bm{r},\bm{p}_k)
&=& D^{(\alpha)}(\bm{r},\bar{\bm{p}}) + D^{(\alpha)}(\bar{\bm{p}},\bm{p}_k)
-\kappa D^{(\alpha)}(\bm{r},\bar{\bm{p}})D^{(\alpha)}(\bar{\bm{p}},\bm{p}_k) \\
\fl &=& D^{(\alpha)}(\bm{r},\bar{\bm{p}}) + D^{(\alpha)}(\bar{\bm{p}},\bm{p}_l)
-\kappa D^{(\alpha)}(\bm{r},\bar{\bm{p}})D^{(\alpha)}(\bar{\bm{p}},\bm{p}_l)
	=D^{(\alpha)}(\bm{r},\bm{p}_l).
\end{eqnarray*}
Hence, ${\cal B}$ is a bisector of $\bm{p}_k$ and $\bm{p}_l$.
The projective equivalence ensures that 
${\cal B}$ is also $\nabla^*$-autoparallel. 

ii) Recall the equality $D^{(\alpha)}(\bm{p},\bm{r})=D^{(-\alpha)}(\bm{r},\bm{p})$ 
and the conformal relation (\ref{conf_div}) 
between $D^{(-\alpha)}$ and $\rho$, then we see that 
${\rm Vor}^{(\alpha)}(\bm{p}_k)={\rm Vor}^{({\rm conf})}(\bm{p}_k)$ 
holds on ${\cal S}^n$, where 
\[
	{\rm Vor}^{({\rm conf})}(\bm{p}_k):=\bigcap_{l \not= k} 
	\{ \bm{p} \in {\cal S}^n | 
	\rho(\bm{p}_k,\bm{p})< \rho(\bm{p}_l,\bm{p}) \}.
\]
Theorem \ref{thm1}, relations (\ref{first_der}) and (\ref{can_div}) 
imply that $\rho(\bm{p}_k,\bm{p})$ 
is represented with the coordinates $(P_i)$ by
\[
	\rho(\bm{p}_k,\bm{p})=\psi^*(\bm{P})
	-\left( \psi^*(\bm{P}_k) 
	+ \sum_{i=1}^n \frac{\partial \psi^*}{\partial P_i} (\bm{P}_k)
	(P_i(\bm{p}) - P_i(\bm{p}_k)) \right),
\]
where $\bm{P}=\bm{P}(\bm{p})$.
Note that a point $(\bm{P},y_k(\bm{P}))$ in ${\cal H}_k$ is expressed by
%represent the hyperplane that is tangent at $(\bm{P}_i,  \psi^*(\bm{P}_i))$,
%where
\[
	y_k(\bm{P}):= \psi^*(\bm{P}_k) 
	+ \sum_{i=1}^n \frac{\partial \psi^*}{\partial P_i} (\bm{P}_k)
	(P_i(\bm{p}) - P_i(\bm{p}_k)).
\]
Hence, we have $\rho(\bm{p}_k,\bm{p})=\psi^*(\bm{P})-y_k(\bm{P})$.
We see, for example, that the bisector on ${\cal E}^n$ for 
$\bm{p}_k$ and $\bm{p}_l$ is represented as a projection of 
${\cal H}_k \cap {\cal H}_l$.
Thus, the statement follows.
\hfill Q.E.D.

\medskip
The figure 2 and 3 show examples of $\alpha$-Voronoi diagrams 
on the simplex of dimension 2.
In these cases, the bisectors are simultaneously $\nabla^{(\alpha)}$- and 
$\nabla^*$-geodesics.

\medskip
%\noindent
{\it Remark 4}:
In \cite{Matsu2} 
Voronoi diagrams for broader class of divergences (contrast functions) 
that are not necessarily associated with any convex potentials are studied 
from more general affine differential geometric points of views.
The construction algorithm is also given there, which is applicable 
if the corresponding affine immersion is explicitly obtained.

On the other hand, the $\alpha$-divergence defined not only on ${\cal S}^n$ 
but on the positive orthant ${\bf R}_+^{n+1}$ can be represented 
as a remainder of the potential $\Psi$ in (\ref{potential}) 
\cite{Amari,AN,Ohara09}.
Hence, the $\alpha$-geometry on  ${\bf R}_+^{n+1}$ is dually flat.
Using this property, $\alpha$-Voronoi diagrams on ${\bf R}_+^{n+1}$ is 
discussed in \cite{NN}.

While both of the above methods require computation of the polyhedrons
in the space of dimension $n+2$, the new one proposed in this paper does 
in the space of dimension $n+1$.
Since the optimal computational time of polyhedrons 
depends on the dimension $d$ by $O(m\log m+m^{\lfloor d/2 \rfloor})$ 
\cite{Chaz}, the new one where $d=n+1$ is slightly better when $n$ is even.
%slightly better.

\medskip

The next proposition is a simple and relevant application of 
escort probabilities. 
Define the {\em $\alpha$-centroid} $\bm{c}^{(\alpha)}$ for given $m$ points 
$\bm{p}_1, \cdots, \bm{p}_m$ on ${\cal S}^n$ by the minimizer of 
the following problem:
\[
	\min_{\bms{p} \in {\cal S}^n} 
		\sum_{k=1}^m D^{(\alpha)}(\bm{p}_k,\bm{p}).
\]
\begin{proposition}
The {\em $\alpha$-centroid} $\bm{c}^{(\alpha)}$ for given $m$ points 
$\bm{p}_1, \cdots, \bm{p}_m$ on ${\cal S}^n$ is represented 
in escort probabilities by the weighted average of conformal factors 
$\lambda(\bm{p}_k)=1/Z_q(\bm{p}_k)$, i.e., 
\[
	P_i(\bm{c}^{(\alpha)})= \frac{1}{\sum_{k=1}^m Z_q(\bm{p}_k)} 
		\sum_{k=1}^m Z_q(\bm{p}_k) P_i(\bm{p}_k), 
		\quad i=1,\cdots,n+1.
\]
\end{proposition}
\noindent
Proof) Let $\theta^i=\theta^i(\bm{p})$. Using (\ref{conf_div}), 
(\ref{can_div}) and the relation 
$D^{(\alpha)}(\bm{p},\bm{r})=D^{(-\alpha)}(\bm{r},\bm{p})$, we have
\[ \fl
	\sum_{k=1}^m D^{(\alpha)}(\bm{p}_k,\bm{p})=\sum_{k=1}^m Z_q(\bm{p}_k)
		\rho(\bm{p},\bm{p}_k)
	=\sum_{k=1}^m Z_q(\bm{p}_k)\{\psi(\bm{\theta})
	+\psi^*(\bm{\eta}(\bm{p}_k))-\sum_{i=1}^n \theta^i \eta_i(\bm{p}_k)\}.
\]
Then the optimality condition is 
\[
	\frac{\partial}{\partial \theta^i} 
	\sum_{k=1}^m D^{(\alpha)}(\bm{p}_k,\bm{p})
	= \sum_{k=1}^m Z_q(\bm{p}_k)(\eta_i-\eta_i(\bm{p}_k))=0, 
	\quad i=1,\cdots, n,
\]
where $\eta_i=\eta_i(\bm{p})$.
Thus, the statement follows from Theorem \ref{thm1} for $i=1,\cdots,n.$
For $i=n+1$ it follows from the fact that the sum of the weights is 
equal to one.
\hfill Q.E.D.

\section{Concluding remarks}
We have considered $\pm 1$-conformal transformations of the 
$\alpha$-geometry and 
obtained dually flat structure $({\cal S}^n,h,\nabla,\nabla^*)$.
Further the potential functions and 
dually flat coordinate systems associated with the structure have been derived.
We see that the escort probability naturally appears 
to play an important role.

\begin{figure}[h]
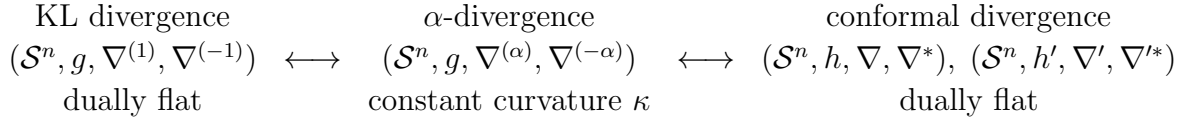

\[ \fl
\begin{array}{ccccc}
	\mbox{KL divergence} &  
	&\mbox{$\alpha$-divergence} & 
	& \mbox{conformal divergence} \\
	({\cal S}^n, g, \nabla^{(1)}, \nabla^{(-1)}) 
	&\longleftrightarrow 
	&({\cal S}^n, g, \nabla^{(\alpha)}, \nabla^{(-\alpha)}) 
%	\stackrel{f}{ \longrightarrow}
%	(f({\cal S}^n) \subset {\bf R}^{n+1}, \tilde \nabla^{(\alpha)},\tilde g
	& \longleftrightarrow 
	&({\cal S}^n,h,\nabla,\nabla^*),\; ({\cal S}^n,h',\nabla',\nabla'^*) \\
	\mbox{dually flat} &  
	&\mbox{constant curvature $\kappa$} & 
	& \mbox{dually flat} \\
\end{array}
\]
\caption{transformations of dualistic structures}
\label{fig2}
\end{figure}

From a viewpoint of contrast functions, 
the geometric structure compatible to the Kullback-Leibler divergence is 
$({\cal S}^n,g,\nabla^{(1)}, \nabla^{(-1)})$, 
where $g$ is the Fisher information and $\nabla^{(\pm 1)}$ are respectively 
the {\em e-connection} and the {\em m-connection}.
Similarly, the $\alpha$-divergence (or the 
Tsallis relative entropy), and the conformal divergence $\rho$ in this note 
correspond to 
$({\cal S}^n, g, \nabla^{(\alpha)}, \nabla^{(-\alpha)})$ and 
$({\cal S}^n,h,\nabla,\nabla^*)$, respectively.
They are summarized in Figure \ref{fig2}.

The physical meaning or essence underlying these transformations would be 
interesting and significant, but is left unclear.
(See recent publications \cite{Ojima,Tanaka} for 
such research directions.)

Finally, we have shown a direct application of the conformal flattening 
to computation of $\alpha$-Voronoi diagrams and $\alpha$-centroids.
Escort probabilities are found to work as a suitable coordinate system 
for the purpose.

%We hope that arguments in this note might be helpful to more 
%profound investigations and understandings of this area
%\cite{Ojima04,Ojima,Tanaka}.

\ack
The first author would like to thank Prof. Tatsuaki Wada for 
helpful comments.
%the anonymous referee who has pointed out several errors 
%in the manuscript and important literature.

\appendix

\section*{Appendix A: 
Statistical manifold and $\alpha$-conformally equivalence}
\setcounter{section}{1}
For details of this appendix see \cite{Lau,Kurose90,Kurose94,Kurose02,Matsu}.
For a torsion-free affine connection $\nabla$ and a pseudo Riemannian 
metric $g$ on a manifold $\mathcal{M}$, the triple $(\mathcal{M},g,\nabla)$ 
is called a {\em statistical manifold} if 
%$\nabla g$ is symmetric.
it admits another torsion-free connection $\nabla^*$ satisfying
\begin{equation}
 Xg(Y,Z)=g(\nabla_X Y, Z)+g(Y, \nabla^*_X Z)
\label{dualrel}
\end{equation}
for arbitrary $X,Y$ and $Z$ in ${\mathcal X}(\mathcal{M})$, where ${\mathcal X}(\mathcal{M})$ is the set of all tangent vector fields on $\mathcal{M}$.
It is known that $(\mathcal{M},g,\nabla)$ is a statistical manifold 
if and only if $\nabla g$ is symmetric, i.e., 
$(\nabla_X g)(Y,Z)$ is symmetric with respect to $X,Y$ and $Z$.
We call $\nabla$ and $\nabla^*$ {\em duals} of each other with respect to $g$, 
and $({\cal M},g,\nabla^*)$ is said the dual statistical manifold 
of $({\cal M},g,\nabla)$.
The triple of a Riemannian metric and a pair of dual connections 
$(g,\nabla,\nabla^*)$ satisfying (\ref{dualrel}) is called a 
{\em dualistic structure} on $\mathcal{M}$.

For $\alpha \in \mathbf{R}$, statistical manifolds 
$(\mathcal{M},g,\nabla)$ and $(\mathcal{M},g',\nabla')$ are said to be
{\em $\alpha$-conformally equivalent} if there exists 
a positive function $\phi$ on $\mathcal{M}$ such that
\begin{eqnarray*}
  g'(X,Y)&=& \phi g(X,Y), \\
 g( \nabla'_X Y,Z)&=&g(\nabla_X Y,Z)-\frac{1+\alpha}{2}d (\ln \phi)(Z)g(X,Y) 
  \\
 && \quad +\frac{1-\alpha}{2}\{d(\ln \phi)(X)g(Y,Z)+d(\ln \phi)(Y)g(X,Z)\}.
%  g'(X,Y)&=&e^\phi g(X,Y), \\
% g( \nabla'_X Y,Z)&=&g(\nabla_X Y,Z)-\frac{1+\alpha}{2}d \phi(Z)g(X,Y) 
%  +\frac{1-\alpha}{2}\{d\phi(X)g(Y,Z)+d\phi(Y)g(X,Z)\}.
%% g( \nabla'_X Y,Z)&=&g(\nabla_X Y,Z)-\frac{1+\alpha}{2}d \phi(Z)g(X,Y) \\
%% && +\frac{1-\alpha}{2}\{d\phi(X)g(Y,Z)+d\phi(Y)g(X,Z)\}.
\end{eqnarray*}
Statistical manifolds $(\mathcal{M},g,\nabla)$ and 
$(\mathcal{M},g',\nabla ')$ are $\alpha$-conformally
equivalent if and only if $(\mathcal{M},g,\nabla^*)$ and 
$(\mathcal{M},g,{\nabla'}^*)$ are 
$-\alpha$-conformally equivalent.

A statistical manifold $(\mathcal{M},g,\nabla)$ is called
{\em $\alpha$-conformally flat} if it is locally
$\alpha$-conformally equivalent to a flat statistical manifold.
Note that $-1$-conformal equivalence implies projective equivalence.
A statistical manifold of dimension greater than three 
has constant curvature if and only if it is $\pm 1$-conformally
flat.

We call a function $\rho$ on ${\cal M}\times {\cal M}$ 
a {\em contrast function} \cite{Eguchi} {\em inducing} 
$(g,\nabla,\nabla^*)$ if it satisfies
\begin{eqnarray}
	\rho(p,p)&=&0,  \quad p \in {\cal M}, \\
	\rho[X|]&=&\rho[|Y]=0,  \\
	g(X,Y)&=&-\rho[X|Y], \label{cont1} \\
	g(\nabla_X Y,Z)&=&-\rho[XY|Z], \quad g(Y,\nabla^*_X Z)=-\rho[Y|XZ],
	\label{cont2}
\end{eqnarray}
where
\[
	\rho[X_1 \cdots X_k|Y_1 \cdots Y_l](p)
	:=(X_1)_p \cdots (X_k)_p(Y_1)_q \cdots (Y_l)_q \rho(p,q)|_{p=q}
\]
for arbitrary $p,q \in {\cal M}$ and $X_i ,Y_j \in {\cal X}({\cal M})$.
If $(\mathcal{M},g,\nabla)$ and 
$(\mathcal{M},g',\nabla ')$ are $1$-conformally equivalent, 
a contrast function $\rho'$ inducing $(g',\nabla ', \nabla'^*)$ 
is represented by $\rho$ inducing $(g,\nabla,\nabla^*)$, as
\[
%	\rho'(p,q)=e^{\phi(q)} \rho(p,q).
	\rho'(p,q)=\phi(q) \rho(p,q).
\]
%where $\rho$ is a contrast function of $(\mathcal{M},g,\nabla)$.

\section*{Appendix B: The proof for the fact that $({\cal S}^n,h,\nabla^*)$ 
is a statistical manifold}
We show that $\nabla^* h$ is symmetric.
By the definition of $-1$-conformally flatness we have
\begin{eqnarray*}
	(\nabla^* _X h)(Y,Z) &=& Xh(Y,Z)-h(\nabla^*_X Y,Z)-h(Y,\nabla^*_X Y) \\
	&=& d\lambda(X)g(Y,Z)+\lambda X g(Y,Z) \\
	&& -\lambda \{g(\nabla^{(\alpha)}_X Y, Z) + d(\ln \lambda)(Y)g(X,Z)
		+d(\ln \lambda)(X)g(Y,Z) \} \\
	&& -\lambda \{g(Y, \nabla^{(\alpha)}_X Z) + d(\ln \lambda)(Z)g(X,Y)
		+d(\ln \lambda)(X)g(Z,Y) \}.
\end{eqnarray*}
Substitute the equality $\lambda d (\ln \lambda) = d \lambda$ into the right-hand side, then it is transformed to
\begin{eqnarray*}
%	(\nabla^* _X h)(Y,Z) 
\fl	&& \lambda \{Xg(Y,Z)-g(\nabla^{(\alpha)}_X Y, Z) 
		-g(Y, \nabla^{(\alpha)}_X Z) \\
\fl		&& \qquad - d(\ln \lambda)(X)g(Y,Z) - d(\ln \lambda)(Y)g(X,Z) 
		- d(\ln \lambda)(Z)g(X,Y) \} \\
\fl	&&= \lambda (\nabla^{(\alpha)}_X g)(Y,Z) 
	-\lambda\{d(\ln \lambda)(X)g(Y,Z) + d(\ln \lambda)(Y)g(X,Z) 
		+ d(\ln \lambda)(Z)g(X,Y)\}.
\end{eqnarray*}
Thus, $\nabla^* h$ is symmetric because 
$({\cal S}^n,g,\nabla^{(\alpha)})$ is a statistical manifold, i.e.,  
$\nabla^{(\alpha)}g$ is symmetric.
Since $\nabla^{(\alpha)}$ is torsion-free, so is $\nabla^*$ by 
the definition of $-1$-conformally flatness.

\end{document}